\documentclass[11pt]{article}
\usepackage[english]{babel}
\usepackage{amsthm}
\usepackage{amsmath}
\usepackage{amsfonts}
\usepackage{amssymb}
\usepackage{makeidx}

\usepackage{graphicx}
\usepackage[normalem]{ulem}
\usepackage{color}
\usepackage[colorlinks=true,linkcolor=blue,citecolor=blue,urlcolor=black]{hyperref}

\usepackage{fancyhdr}
\providecommand{\ams}[1]{\textit{AMS 2010 Subject Classification.} #1}

\providecommand{\keywords}[1]{\textit{Keywords.} #1}

\newcommand{\beq}{\begin{eqnarray}}
\newcommand{\eeq}{\end{eqnarray}}
\newcommand{\beqq}{\begin{eqnarray*}}
	\newcommand{\eeqq}{\end{eqnarray*}}
\def\:{:\,}

\newcommand{\ph}{\varphi}

\def\mR{\mathbb{R}}

\def\tX{{\tilde X}}
\def\tx{{\tilde x}}
\def\real{\mR}

\def\definedas{\stackrel{\Delta}{=}}
\def\bc{\begin{center}}
\def\ec{\end{center}}

\def\definedas{\stackrel{\Delta}{=}}

\newcommand{\cX}{{\mathcal X}}

\newcommand{\cN}{{\mathcal N}}

\newcommand{\cL}{{\mathcal L}}

\def\1{\mathbf{1}}

\begin{document}

\title{Modeling  and replicating statistical topology, and evidence for CMB non-homogeneity}



\author{Robert J.\ Adler, Sarit Agami, and Pratyush Pranav\\   \\
Andrew and Erna Viterbi Faculty of  Electrical Engineering\\
Technion -- Israel Institute of Technology}

\maketitle

\begin{abstract}
{\footnotesize
Under the banner of `Big Data', the detection and classification of  structure in extremely large, high dimensional, data sets, is,  one of the central statistical challenges of  our times. Among the most intriguing approaches to this challenge is `TDA', or `Topological Data Analysis', one of the primary aims of which is providing non-metric, but topologically informative,  pre-analyses of data sets which make later, more quantitative analyses feasible. While TDA rests on strong mathematical foundations from Topology, in applications it has faced challenges due to an inability to handle issues of statistical reliability and robustness and, most importantly,  in an inability to make scientific claims with verifiable levels of statistical confidence. We propose a  methodology for the parametric representation, estimation, and replication of persistence diagrams, the main diagnostic tool of TDA. The  power of the methodology lies in the fact that even if only one persistence diagram is available for analysis -- the typical  case for big data applications --   replications can be generated to allow for conventional  statistical hypothesis testing. The methodology is conceptually simple and computationally practical, and provides a broadly effective statistical procedure for  persistence diagram TDA analysis.  We demonstrate the basic ideas on a  toy example, and the power of the approach in a novel  and revealing analysis of CMB non-homogeneity.}
\end{abstract}

\ams{ Primary: 60G55, 
62H12,  	
62H15.  	
 \\
		Secondary: 62G05, 
				85A35. 
	}
	

%
%
%


\section{Introduction}

As a consequence of the current explosion in size, complexity, and  dimensionality of data sets, there has been a growing need for the development of powerful but concise summary statistics and visualisation methods that facilitate  understanding and decision making. A singularly novel approach, which has been particularly promising  in areas as widespread as biology and medicine   \cite{PNAS1,PNAS2,PNAS3}, neurophysiology, \cite{CURTO}, cosmology \cite{sousbie2011persistent,PEWVKJ},
and materials science \cite{Yasu}, has been via the application of the powerful, and rather abstract, concepts of Algebraic Topology to develop what generally now falls under the label of `Topological Data Analysis', or TDA.
While approaching complex data from a topological viewpoint is not entirely new -- it underlies Tukey's `Exploratory Data Analysis' of the 1960's \cite{Tukey62} and the more recent approach by Friston and coworkers to brain imaging data \cite{Friston10}  -- TDA differs from all its forebears in its sophisticated exploitation of  recent developments in Computational Topology. In particular, much of TDA has become almost synonymous with an analysis based on some version of  persistent homology  \cite{EdelsShortCourse,EdelsHarerSurvey,EdelsHarerBook}, represented visually  as barcodes, persistence diagrams, or related  representations  \cite{Carlsson-review,CarlssonReview,GHRIST-BARCODES,ghrist2014elementary,WassermanReview}.

With relatively few exceptions, notably \cite{chazal,fasy,robinson,OmerSayan,bubenik,RT}, TDA has not employed statistical methodology as part of its approach, and, as a consequence, has typically been unable to associate clearly defined levels of statistical significance to its discoveries. While there may be a variety of reasons for this, one of the main obstacles to doing so is that  the mathematical challenges involved in computing the statistical distributions of topological quantifiers have so far proven to be intractable. This is despite the fact that the measure-theoretic issues involved in defining probability spaces for objects related to persistent homology have indeed been solved; e.g.\ \cite{TurnerFrechet2,TurnerFrechet1}.

One approach adopted by \cite{chazal,fasy,robinson} and others to circumvent these difficulties has been to reduce persistence diagrams to a single test statistic, often related to bottleneck norms, and then to adopt standard statistical resampling methods to analyse this statistic. If multiple diagrams are available, then the resampling can be done  on them.  However, since TDA is typically used in areas of very large data sets, the availability of replicates is rare, and consequently  this approach is impracticable  in most applications. An alternative approach is to (sub)sample points from the persistence diagram, and compute statistics on the subsamples. The problem with this approach, however, is that the true random object here is the full persistence diagram, and it thus it is often unclear what is the precise meaning of the statistics produced this way.

We introduce a new approach, based  on generating a sequence of persistence diagrams which has similar statistical properties to those of the one generated by the data. While novel, the ideas underlying the method are not difficult, and follow a number of clearly defined stages. Firstly, a parametric model is adopted that is sufficiently  flexible to model an extremely wide class of persistence diagrams. The model we use is a class of Gibbs distributions, since these have a long history of success in modelling point sets, (\cite{BCG} and its bibliography)  and, in the final analysis, this is what a persistence diagram is. Having estimated parameters, we then exploit the fact that Gibbs distributions lend themselves to simulation by Markov chain Monte Carlo (MCMC) methods, and we exploit MCMC to produce a simulated  sequence of diagrams. Since the underlying raison d'\^etre of this approach is that  persistence diagrams provide an excellent summary of topology, and statistics computed off the diagrams themselves furnish even more succinct summaries, we call this procedure {\it  Replicating Statistical Topology}, or {\it RST}, and its introduction and the descriptions of its implementation  are the main contributions of the paper.
 We believe that these ideas will provide a significant contribution towards putting TDA on a more solid statistical footing. To support this we shall treat one toy example, showing that the technique works as predicted, and then study the fascinating and important topic of
 non-homogeneities in the CMB (Cosmic Microwave Background) radiation using an RST approach.

\section{TDA and persistence diagrams}


As homology is an algebraic method for describing topological features of shapes and functions, so persistent homology is an extension of this method for both  enriching these descriptions and for describing how topology undergoes changes. We shall use it to describe the upper-level sets of real-valued functions $f$ defined over a space $\cX$; viz.\ sets of the form $A_u=\{x\in \cX\: f(x)\geq u\}$. In basic homology, the topology of each $A_u$ is often summarized by its Betti numbers, $\beta_k$, $k=0,\dots \text{dim}(\cX)$. The first of these, $\beta_0$, counts the number of connected components in $A_u$, and, roughly speaking, the remaining $\beta_k$ count the number of $(k+1)$-dimensional `holes' in $A_u$. Persistent homology goes further, and keeps track of how the homology, including the Betti numbers,  changes as a function of the level $u$, giving a richer, more dynamic view of topology.

\begin{figure}[h]
\bc
\includegraphics[width=1.45in, height=1.45in]{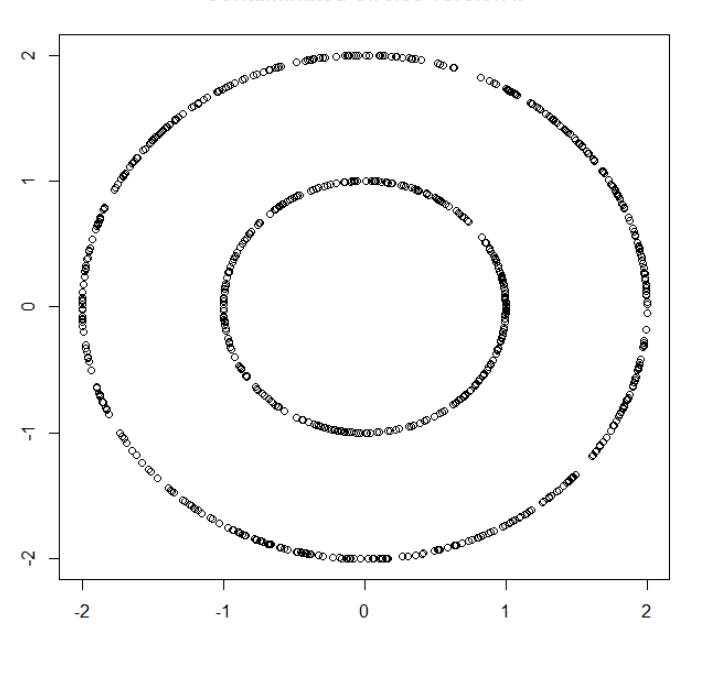} \hskip0.01truein
\includegraphics[width=1.45in, height=1.45in]{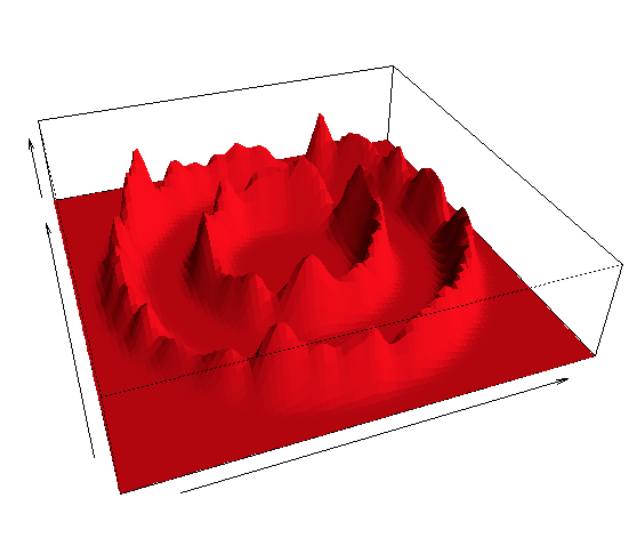} \hskip0.01truein
\includegraphics[width=1.45in, height=1.45in]{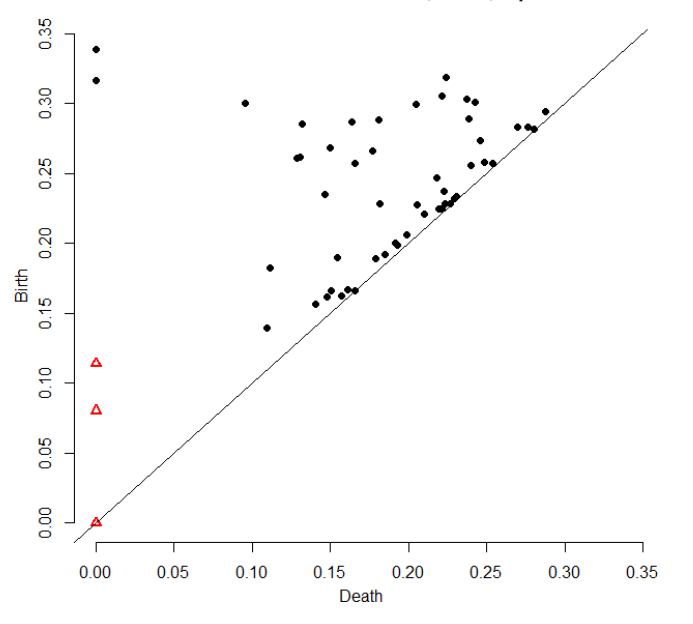}
\ec
\caption{\footnotesize
 A random sample from two circles, 500 points from the larger circle and 300 from the smaller one, along with a kernel density estimate
and the persistence diagram for its upper level sets. The black circles are the $H_0$ persistence points, while the red triangles are the $H_1$ points.
}
\label{fig:twocircles}
\end{figure}

Today persistent homology is undeniably the most popular tool in the burgeoning area of TDA, one of the main reasons for which is the fact  that it is easily visualised via barcode diagrams. Each bar in such a diagram is an interval that  starts (is `born' at) a level $b$, at which a new aspect of the homology of  $A_u$ appears, and ends (`dies' at) a level $d$, as this aspect disappears. A mathematically equivalent, but visually distinct representation, of barcodes is as persistence diagrams, henceforth PDs, of the points $(b,d)$.  We shall assume that the reader has some familiarity with these concepts, but look at an instructive, and  easy, example in Fig.\ \ref{fig:twocircles}, needed later.

At the  left, we see a sample $\tx_N =\{x_1,\dots,x_N\}$ of $N=800$  points from two circles,  of diameters  4 and 2. A random sample of  500 points were chosen at random from the larger circle, and 300 from the smaller one. To its right,  we see the corresponding kernel density estimate, defined by
\beq
\hat f_N(p) \ = \ \frac{1}{ N} \sum_{i=1}^{N} \frac{1}{2\pi\delta^2}  e^{{-\|p-x_i\|^2}/{2\delta^2}},
\eeq
 where
 $\delta >0$ is a bandwidth parameter, the precise value of which is not important for now.

At the far right
we have the corresponding persistence diagram of the upper level set filtration of $\hat f_N$, with the black circles indicating $H_0$ (zero-th homology) persistence and the red triangles corresponding to $H_1$, in both cases trying to capture the underlying homology of the two circles.  As described above, each point  in the diagram is a `birth-death' pair $(b,d)$. The accepted paradigm of TDA is that points in the PD `far away' from the diagonal $b=d$ are meaningful, while points close to the diagonal, which represent short lived topological phenomena, are not. Thus, since we know that the upper level sets of $\hat f_N$ are characterized by having two main components, each of which contains a single 1-cycle (hole) we expect to see  two black circles and two red triangles somewhat isolated from the other points in the diagram.  This is in fact the case.


While the  PD in Fig.\  \ref{fig:twocircles} performs as expected, and it is easy to identify the points that, a priori, we knew had to be there, there are many other points in the diagrams, and were we not in the situation of knowing ahead of time which points were `really' significant, it would not have been clear how to discount the additional points. We shall see later how to do this, but first must describe the general approach.

\section{Gibbs measures for persistence diagrams}
\label{sec:Gibbs}

Given a finite collection $\tX_N=\{X_1,\dots,X_N\}$ of continuous random variables, with joint probability density $\ph_\Theta(\tx_N)$,
dependent on a multi-dimensional parameter $\Theta$, we say that
$\ph_\Theta$ is a Gibbs distribution if it is written in the form
\beq
\ph_\Theta(\tx_N) = \frac{1}{Z_\Theta} \exp ( - H_\Theta (\tx_N)),
\label{eq:Gibbs}
\eeq
where the `Hamiltonian' $H_\Theta\:\real^N\to\real$
describes the `energy' of configurations $\tx_N$. The normalisation $Z_\Theta$, actually a function of $\Theta$, is known as the partition function, and is infamously hard to evaluate. All this is standard fare \cite{Georgii}. What is new is that we shall use Gibbs distributions to provide a model for PDs  that look like those in Fig.\ \ref{fig:twocircles}. (There is another important family of PDs that arise from the construction of simplicial complexes over point sets, and these, at least for their $H_0$ diagrams, have all points with birth times identically zero. These are much easier to analyze, since they are effectively one dimensional, and we shall treat them in a separate publication.)

As in all applications of Gibbs distributions, success depends on an appropriate choice for the energy function. Here is a way to do it for
$\tx_N$ a set of $N$ points in a subset $\cX$ of $\real^2$.
Firstly, for  $x\in\cX$ and, for $k, \delta>0$ let
$\cN_{\delta,k}(x)$ be the collection of the $k$ nearest neighbours of $x$ in $\tx_N$, assuming that all are of distance no greater than $\delta$ from $x$.  If $k$ such points do not exist, then we take $\cN_{\delta,k}(x)=\emptyset$.

Assign a weight, or total length to each such cluster, given by the sum of distances to its reference point, viz.
\beqq
\mathcal L_{\delta,k}(x)   \equiv \mathcal L_{\delta,k}(x\, |\,  \cN_{\delta,k}(x) )   =   \sum_{y\in\cN_{\delta,k}(x)}  \|x-y\|.
\eeqq
The weight, or total length, of all  $(k+1,\delta)$-clusters is then
\beqq
\mathcal L_{\delta,k}   \equiv  \mathcal L_{\delta,k}(\tx_N) =  \sum_{x\in\tX_N} \mathcal L_{\delta,k}(x).
\eeqq

Now, for $x\in\cX$ write $x=(x^{(1)},x^{(2)})$,   and
\beqq
\sigma_H^2 =  \sum_{x\in\tx_N} \big(x^{(1)} - \bar x^{(1)}\big)^2,\ \ \
\sigma_V^2 = \sum_{x\in\tx_N} \big(x^{(2)} \big)^2,
\eeqq
where $\bar x^{(1)}=N^{-1} \sum_{i=1}^N x_i^{(1)}$.

Thinking now, for reasons to become clear soon, of $\cX=\real\times\real_+$, we  define a Hamiltonian, at effective interaction distance $\delta$,  up to cluster size $K\geq 0$, and with interaction parameter $\Theta=(\theta_H,\theta_V,\theta_1,\dots,\theta_K)$, as
\beq
\label{eq:hamiltonian}
H_{\delta,\Theta}^K(\tx_N)
= \theta_H \sigma^2_H
+\theta_V  \sigma^2_V
+\sum_{k=1}^K  \theta_k \mathcal L_{\delta,k}(\tx_N).
\eeq
The parameters here all have a very clear meaning. The horizontal spread of the points in $\tx_N$ is controlled by $\sigma^2_H$, the vertical spread by $\sigma^2_V$, and each $\theta_k$ controls the probability of clusters of size $k+1$, with $\theta_k<0$ favouring such clusters, and $\theta_k>0$ lowering their probabilities.

Now, given a PD $\widetilde B =\{(b_i,d_i)\}_{i=1}^N$,  define a new set of $N$ points $\tx_N =\{x_i\}_{i=1}^N$, with $x^{(1)}_i=b_i$ and
 $x^{(2)}_i=d_i-b_i$. This (invertible) transformation has the effect of moving the points in Fig.\ \ref{fig:twocircles} downwards, so that the diagonal line projects onto the horizontal axis, but still leaves a visually informative diagram, which we shall call the projected PD, or PPD. The statistical model we take for  PPDs is a  Gibbs distribution \eqref{eq:Gibbs} with Hamiltonian \eqref{eq:hamiltonian}.

While this may seem a rather arbitrary form for the distribution of a PPD, there are two facts justifying it. The first is the trivial observation that {\it any} multivariate distribution can be written in the form \eqref{eq:Gibbs}, simply by taking $H_\Theta \equiv -\ln (\ph_\Theta)$ and $Z_\Theta=1$. The second is that, having done this, we would like to take $H$ from a rich enough family of functions to come close to spanning all `reasonable' functions on PPDs. However, we know from  \cite{GunnarRing}
that the ring of algebraic functions on the space of PPDs is spanned by monomials of the form
$(x^{(1)}_1-x^{(2)}_1)^{m_1}(x^{(2)})^{n_1}\cdots (x^{(1)}_l-x^{(2)}_l)^{m_l}(x^{(2)}_l)^{n_l}$, for which $n_i>0$ implies $m_i>0$, and  functions of the form  \eqref{eq:hamiltonian} form a rich subset of these monomials. Furthermore, `cluster expansions' of this form have been successfully  employed in Statistical Mechanics for the best part of a century as a basic approximation tool in the study of Gibbs distributions.

The determination of $\delta$ depends on the number of the points $N$ on the persistence diagram, and on the spread of the points. In practice, theoretical results (cf.\ the reviews \cite{KahleR,BobK}) suggest taking  $\delta$ of the form
  \beq
  \label{eq:delta}
 \delta =\frac{\delta^{*}}{N^{\alpha_{k,d}}}
 \max\left(  \max |x^{(1)}_i - x^{(1)}_j|,\    \max |x^{(2)}_i - x^{(2)}_j| \right),
  \eeq
where $\alpha_{0,d}=1/d$, $\alpha_{k,d}={k/(k+1)d}$, for $k\geq 1$, $d$ is the dimension of the data underlying the PD,
 and $\delta^{*} $ is a data independent tuning parameter.
 The terms inside the brackets in \eqref{eq:delta} scale for the order of magnitude of the data, which is unimportant topologically.
  (For  cases for which $d$ is unknown, setting $d=2$ seems to work in practice, merely leading to larger than usual values of $\delta^{*}$, as does ignoring the fine structure of $\alpha_{k,d}$ and taking it to be $1/\sqrt{N}$, as a global default.)

\subsection{Pseudolikelihood}
Given $H_\Theta$ as a parametric model, we now turn to the estimation problem. Unfortunately, estimation of the parameters by a method such as direct maximum likelihood is precluded by the fact that we neither have an analytic form for  $Z_\Theta$, nor is there any way to compute it numerically in any reasonable time frame.

The standard way around this problem, which we adopt, is the pseudolikelihood approach  \cite{Besag,chalmond}. This originated  in the context of  point cloud data with spatial dependence, which is, essentially, a description of a PD.
In particular, it exploits the inherent spatial Markovianess of a Gibbs distribution to replace the overall probability of, in our case, a random PPD $\tX_N$ by the pseudolikelihood
\beq
\label{eq:pseudo}
L^K_{\delta,\Theta}(\tx_N)  \definedas
\prod _{x\in\tx_N}  f_\Theta \left(x\big|    N_{\delta,K}(x) \right),
\eeq
 where the conditional, local, densities $f_\Theta\left(x\big|    N_{\delta,K}(x) \right)$
 are given by
 \beq
  \frac{
 \exp \left(-H^K_{\delta,\Theta}\left(x\big|   N_{\delta,K}(x) \right)\right)
  }{
\int _{\real}\int _{\real_+} \exp \left(-H^K_{\delta,\Theta}\left(z\big|   N_{\delta,K}(x) \right)\right)\,dz^{(1)}dz^{(2)},
}
\label{eq:conditionalham}
\eeq
and  the conditional, Hamiltonians $H^K_{\delta,\Theta}\left(x\big|   N_{\delta,K}(x)\right)$ by
\beqq
\theta_H\left[x^{(1)}-\bar x^{(1)}\right]^2+\theta_V(x^{(2)})^2 +\sum_{k=1}^K\theta_k\cL_{\delta,k}\left(x\big| N_{\delta,K}(x)\right).
\eeqq

\subsection{Model specification and parameter estimation}
While it might be expected that PPDs originating from different areas might require quite different models, we have found, in all the examples that we tried, that taking $K=2$ in \eqref{eq:hamiltonian} -- so that the largest cluster size was 3 -- was both efficient and sufficient. If a lower $K$ was appropriate, then the estimation procedure described above estimated the higher order parameters $\theta_k$ as close to zero. In this case, using standard, automated statistical  procedures such as   AIC, BIC, etc.\ (cf.\ \cite{burnham}) we often deleted  the corresponding clusters from the Hamiltonian. Overall, we found the procedure not to be sensitive to either these small parameters or the specific procedure adopted for deleting them.  After considerable experimentation, we found that working with all parameters appearing when $K=2$, regardless of their absolute value, was the easiest thing to do. We also found that taking $K>2$  did little to improve the simulation procedure, and typically led to  manifestations of overfitting.

\section{RST and MCMC}

We refer the reader to \cite{RobertCasella,Handbook} for technical background to this section, in which we  describe a standard Metropolis-Hastings MCMC approach to replicating PDs.

Given a pseudolikelihood as in the previous section (with known or  estimated parameters), generating simulated replications of the associated point set  via MCMC is not hard, but first we need some definitions.

Firstly, given a $\tx_N$,  take
$q(\cdot |\tx_N)$ to be the  folded Gaussian density  on $\real^2$ with mean vector and covariance matrix identical to the empirical mean and covariance of the points in $\tx_N$.
Next, for two points $x,x^*\in\cX $  define an `acceptance probability', according to which we will replace $x\in\tx_N$ by $x^*$, thus giving the updated PPD $\tx_N^*$, as
\beqq
\rho \left(x, x^*\right)
= \min \left\{1,\frac{f_\Theta\left(x^*| N_{\delta,K}(x)\right) \cdot q(x|\tx^*_N)}{f_\Theta \left(x|  N_{\delta,K}(x)\right)  \cdot q(x^*|\tx_N)}\right\}.
\eeqq
(Note that  integration in the denominator of $f_\Theta$ in \eqref{eq:conditionalham} depends on $x$ only through its neighbourhood, and so, due to cancellation in the ratio, does not enter into the computation of $\rho \left(x, x^*\right)$. This, of course, is what makes MCMC for pseudolikelihoods so much more computationally feasible than for full likelihood models.)

The basic step of the algorithm, which describes the update of the point set $\tx_N=\left( x_1,\dots,x_N\right)$, is then Algorithm 1.

\begin{figure}[h]
\bc
\includegraphics[scale=0.25]{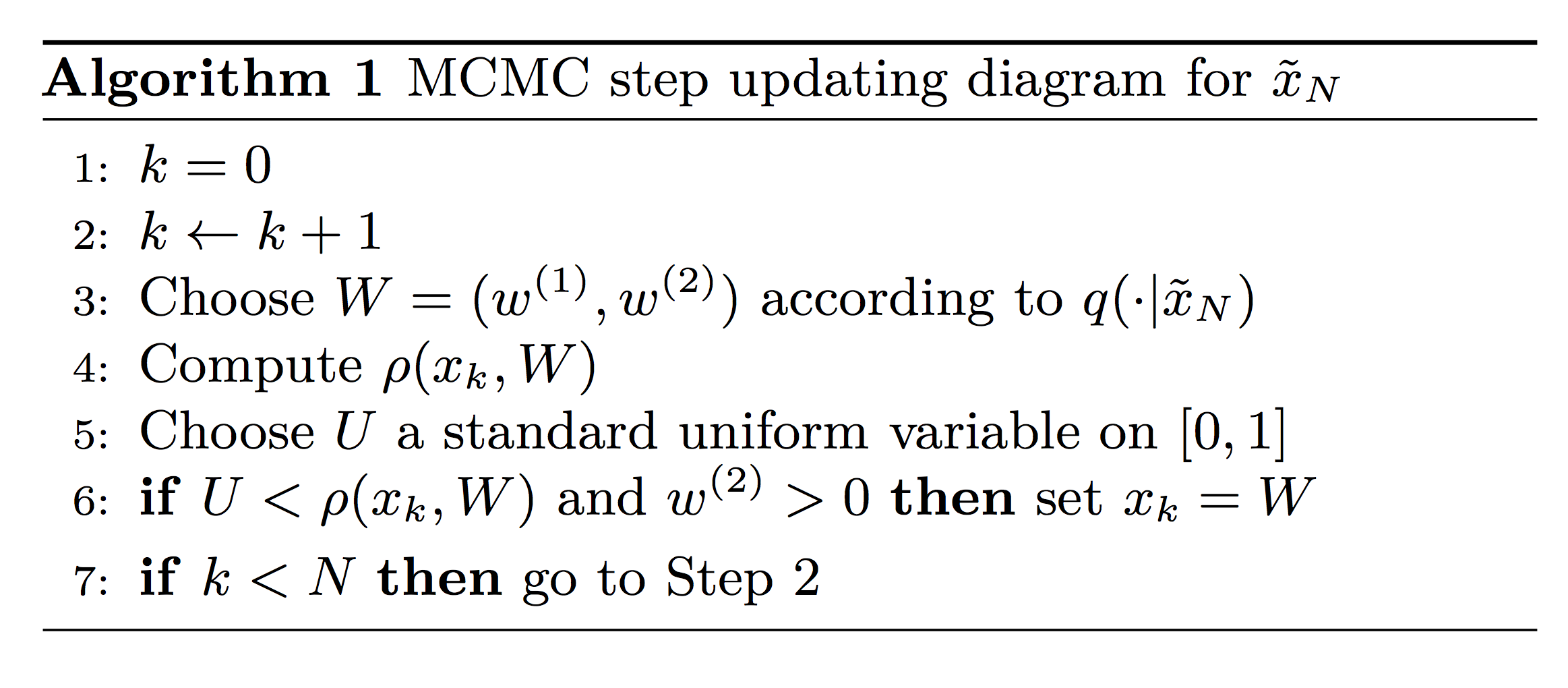}
\label{fig:CMBH}
\ec
\end{figure}


In order to obtain a total of $N$ approximately independent PPD's, we adopt a procedure dependent on three parameters, $n_b$, $n_r$ and $n_R$, as follows:
 Starting with the original PPD,   run Algorithm 1 for a burn in period. Following this, and starting with the final PPD from the burn in,  run the algorithm  a further $n_b$ times, choosing the last output of this block of  $n_b$-th iterations as the first simulated PPD. Repeat this procedure $n_r$ times, each time starting the algorithm with the most recently simulated PPD; viz.\ the output of the previous block.  Finally, replicate this entire procedure $n_R$ times, giving a total of
 $n= n_r\times n_R$ simulated PPDs. For a given $n$, increasing $n_R$ at the expense of the other parameters gives a collection of PPDs more closely related to the original one. Increasing $n_b$ reduces the dependencies between the simulated  PPD's, and so on.

Given the   collection of $n$ simulated PPDs, we convert each PPD back to a regular persistence diagram with the mapping $(x,y)\to (x,x+y)=(b,d)$ of its component points, and write  $\mathcal S({\widetilde B}) = \{\widehat B_1,\dots,\widehat B_n\}$ for the resulting collection of simulated PDs generated from $\widetilde B$.   These form the first level output of the RST procedure.

The  higher levels are  very much  driven by the specific application, but the basic idea is to compute simpler, real or vector valued statistics off the simulated PDs, $\widehat B_i$,  and take their empirical distribution as an estimate of the true, underlying, distribution of the statistic. The same statistic, computed off the original PD $B$, can then be tested for statistical significance against this empirical distribution in standard fashion. This is best described by a simple example.

\section{Examples}
\label{sec:examples}

\subsection{Example 1: Two circles}
As a simple (but representative) test case, we take the random sample from the  two circles in Fig.\ \ref{fig:twocircles}. Note firstly that while there are quite a few (black, circular) points corresponding to the $H_0$ homology, there are only three (red, triangular) ones for $H_1$. Furthermore, the $H_1$ points are all closer to the diagonal boundary, and so less prominent. (These are common phenomena for barcodes, and have been addressed theoretically in a number of studies (e.g.\ \cite{BKS}).)  As a consequence, the  RST procedure will not work for the $H_1$ points in this particular example. However, we do not know of, nor can imagine, any statistical procedure that can reach a meaningful conclusion based on so few points. (Note that  procedures such as those described in \cite{chazal,fasy} require some form of replication, usually via a bootstrapping approach, of the {\it original} data set. This is precisely what we are trying to avoid.) On the other hand, a homology which has at most three generators is small enough to be investigated ad hoc, and statistical procedures are hardly needed.

However, there are certainly  enough $H_0$ points in Fig.\ \ref{fig:twocircles} to fit a spatial model to them. Before we do this, note that there are two points (at the top left) that we know to be significant, since we know, a priori, that the data comes from points on two circles. However there are a number of other points far away from the diagonal, and, were we ignorant of the true situation, it would not be clear as to whether they are significant or not.

%
%

Adopting the approach of RST described above, and working only with the $H_0$ PD, we estimated the parameters for a Gibbs distribution for the model with pseudolikelihood \eqref{eq:pseudo}, taking $K=2$. For three different scenarios, we  generated  1,000 simulated PDs from this model, each with a burn in period of 1,000 iterations and with  $(n_b,n_r,n_R)$ given by (500,20,50), (500,40,25),  or (500,100,10).

Using these three sets of simulations, we computed a number of statistics, but report on only one set here. Perhaps the most natural statistics to look at are the order statistics of the distances of the points in the PD to the diagonal. Given the  points $(b_i,d_i)$,
$i=1,\dots,N$, of the PD, these are $T_j$, the $j$-th largest among the differences $|d_i-b_i |$, $j=1,\dots,N$.  Empirical  distributions of the  order statistics are then trivial to derive from the simulations of the PDs, and  the order statistics calculated off the original PD can be compared to these. The results, for all three scenarios, showed that  $T_1$  and $T_2$ were highly significant (the largest $p$-value
reached in any of the six cases was 0.003). The $p$-values for  $T_3$ were all in the range $(0.036,\, 0.041)$, and so while $T_3$  would be considered significant at the standard 5\% level this is not the  case at  even a 3\% level. In none of the three scenarios was $T_4$ significant, with  $p$-values in the range  (0.13,\, 0.15).

For comparison, we also undertook an analysis using the  bootstrap based techniques of \cite{chazal}, generating  (confidence) bands
above the diagonal. Points lying outside these bands are considered significant.  Both the $H_0$ and $H_1$ bands included all but one point, (and that only marginally) indicating an underlying set topologically equivalent to a single circle, but (markedly) missing the second circle. A similar analysis, using the related  techniques in \cite{fasy},  identified one $H_0$ point but no $H_1$ points at all.

In summary, blindly applying RST to generate simulated PDs, and taking the simplest of all statistics, showed (correctly) strong statistical evidence for two connected components in the topological space (two circles) which generated the PD, with borderline (but misleading) evidence for a third component. Despite the fact that the PD has a number of points far from the diagonal, and quite close to the third furthest point (see Fig.\ \ref{fig:twocircles}) these were (correctly) considered statistically insignificant.

Thus, in this toy example, with the simplest of statistical quantifiers, RST works as hoped, and competes more than favorably with existing methodology.

%
%
%
%

\subsection{Example 2:  CMB non-homogeneity}

Current cosmological theory describes  a phase of rapid
inflation in the primordial universe roughly $10^{-35}$ seconds after its birth. Spontaneous quantum fluctuations in what was then a high energy, uniform, pseudo-vacuum universe,    resulted in minute  perturbations in its density field. Eventually, aided by gravitational amplification, these  fluctuations led to the complicated, inhomogeneous structure of the Cosmic Web of planets, stars, galaxies, etc. which make up today's universe.

The Cosmic Microwave Background (CMB) is the thermal radiation, generated as the universe cooled, some 300,000 years after the hypothesised Big Bang. Amazingly, it is  measurable still today, and since the temperature
fluctuations in the observable CMB follow the pattern of
the quantum perturbations from the inflationary era, it enables the mapping of the fluctuations in the distribution of matter in the early Universe.

\begin{figure}[h]
\bc
\includegraphics[scale=0.15]{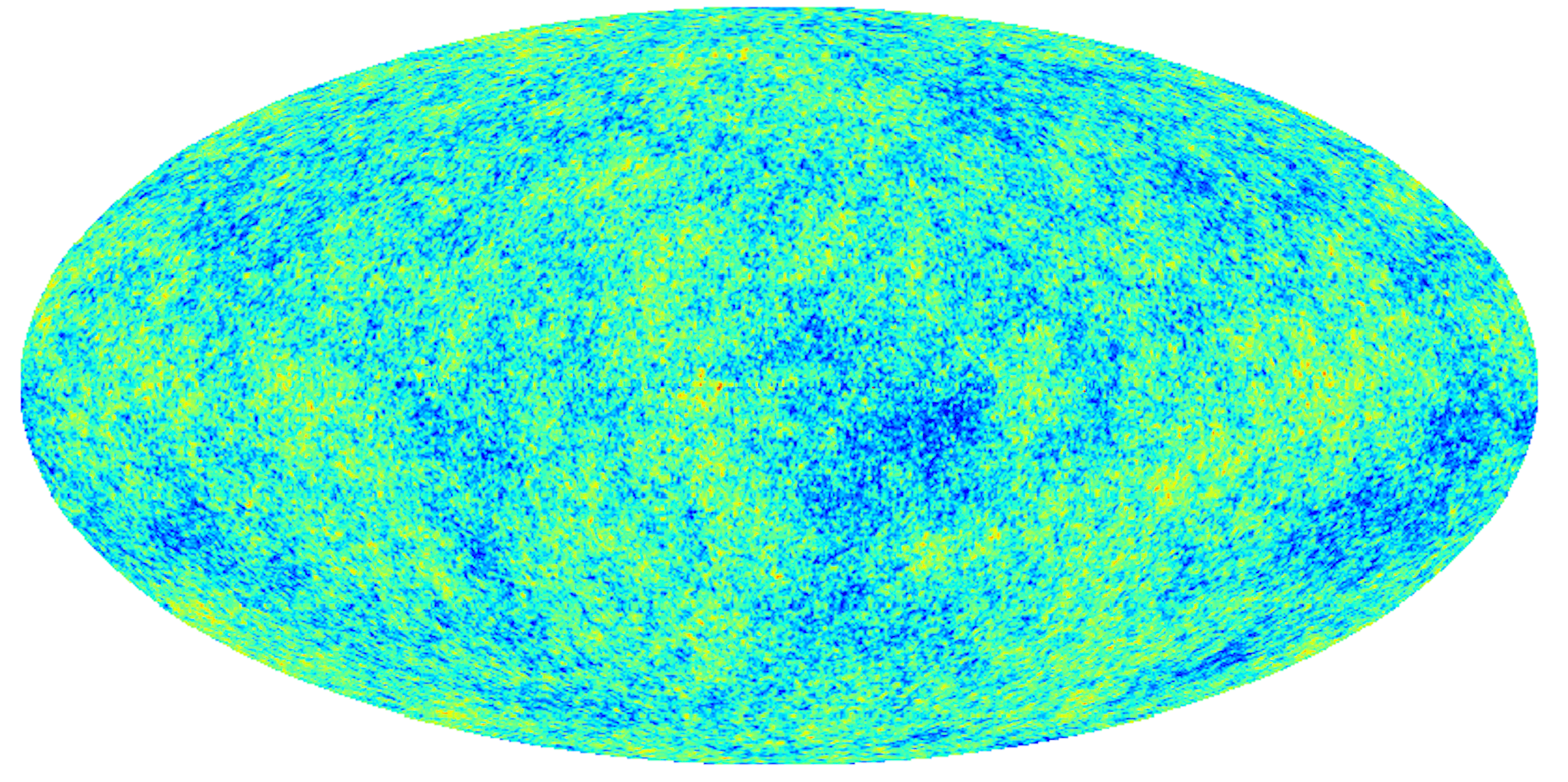}
\ec
\caption{\footnotesize{A reconstructed version of CMB data from the Planck experiment, created using the Commander-Rule technique, seen  in two dimensional, Mollweide projection.  }}
\label{fig:Planck}
\end{figure}

CMB data is directional, measuring  fluctuations in radiation at a number of different frequencies,  coming into a satellite from different directions.  The first, satellite based, detailed measurement of the CMB was carried out by the Cosmic Microwave Background Explorer (COBE) probe  in the early 1990's, followed a decade later by the  Wilkinson Microwave Anisotropy Probe (WMAP).  Most recently, the high precision Planck mission was launched, which measured the temperature anisotropies of the CMB to an accuracy of $10^{-5}$ degrees. It measures the CMB temperature anisotropies at 7 different frequency bands, at a resolution of 5' (i.e. 5 arc-minutes, or  5 sixtieth of a degree), representing the most detailed and precise measurement of the CMB temperature anisotropies till date. Common to all of these, however, is that each CMB measurement is that of a function on a sphere, as in Fig.\ \ref{fig:Planck}.

There are many mathematical  models for the CMB, the most common being that it is a realization of a homogeneous and isotropic
Gaussian random field
\cite{smoot1992,bennett2003,Spergel2007,Komatsu2011,Planck2015cosmoparams}.  Both the assumptions of Gaussianity and homogeneity have been challenged recently, from both theoretical and empirical viewpoints  \cite{Eriksen,Park2004},   and is the issue of homogeneity that we wish to address now, using PDs and Gibbs models. (The issue of Gaussianity is addressed in far more detail,  using topological methods, albeit different to the ones that we are using here, in \cite{Pratyush}, and with geometrical  methods in  \cite{PlanckStatistics,Buchert}.)

In order to test homogeneity, we first cut out a ring around the equator of $\pm 30^\circ$, leaving  `northern' and `southern' $60^\circ$ spherical caps of data. The reason behind ignoring the central ring is that much of  the `data' here is not from actual observations, which are unavailable due to confounding effects such as the Milky Way, but is a `reconstruction' using one of a variety of techniques
\cite{Planck}. Since all of these techniques are based on both Gaussian and homogeneity assumptions, the central ring should not show any  deviation from the assumptions. Our aim is  to test whether or not the CMB in the two caps can be assumed to be realizations of the same stochastic process.

The next step is to generate 8 smoothed version of the CMB in each cap, which we do with 8 different Gaussian kernels, with full width half maximum
300', 180', 120', 90', 60', 40', 20' and 10'. The highest level of smoothing (300') suppresses most of the fine scale variation seen in Fig.\ \ref{fig:Planck}, while the 10' level leads to no visually distinguishable difference. For each such smoothing, we produce PDs generated by the upper level set filtration, for both $H_0$ and $H_1$, leading to a total of $32 = 8\times 2\times 2$ PDs. Although the aims there are different, details of the numerical procedure can be found  in
\cite{PEWVKJ}, and  an example of two PDs is given in Fig.\ \ref{fig:CMBH}.
\begin{figure}[h]
\bc
\includegraphics[width=2.2in, height=2.2in]{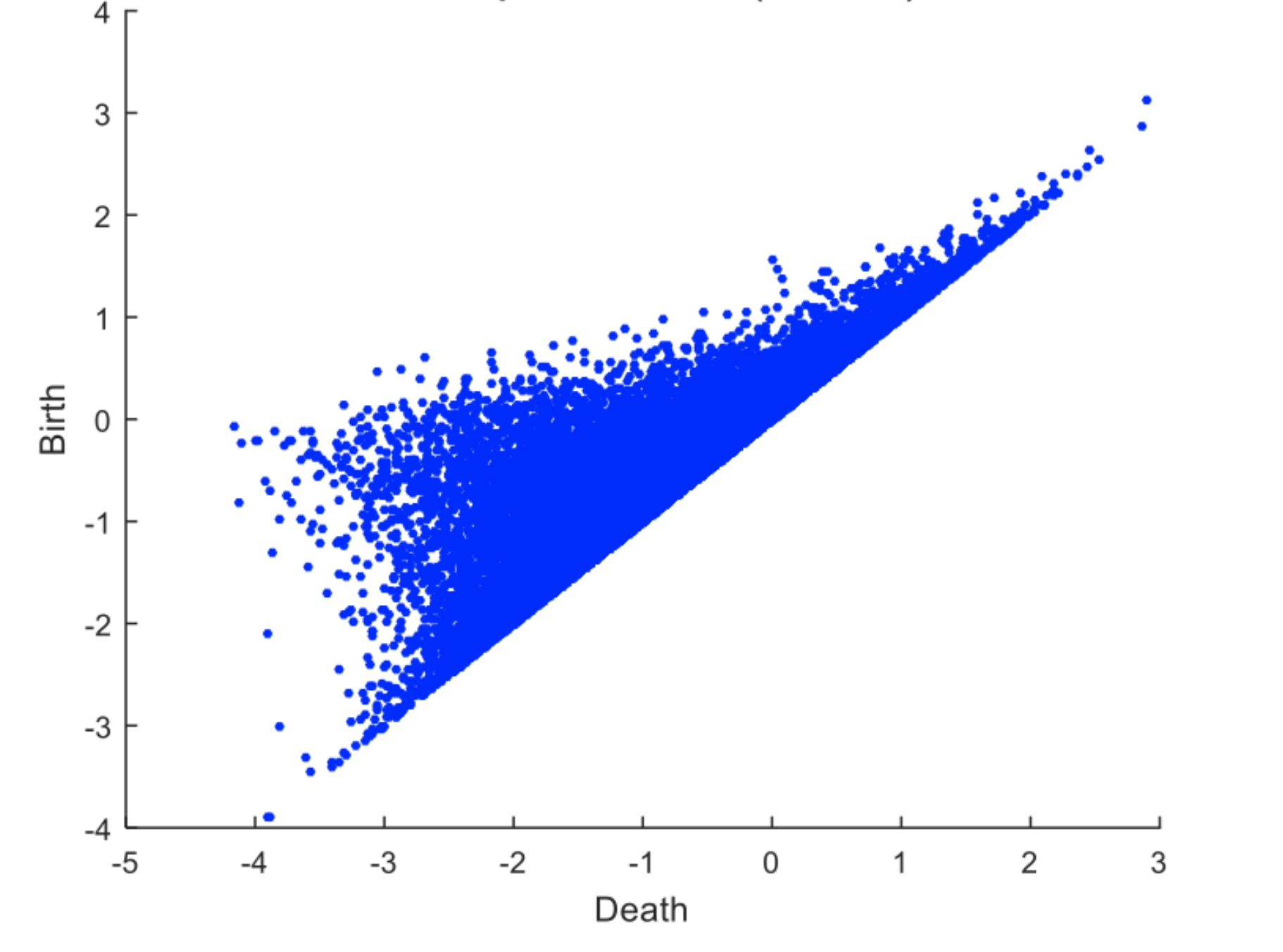} \hskip0.1truein
\includegraphics[width=2.2in, height=2.2in]{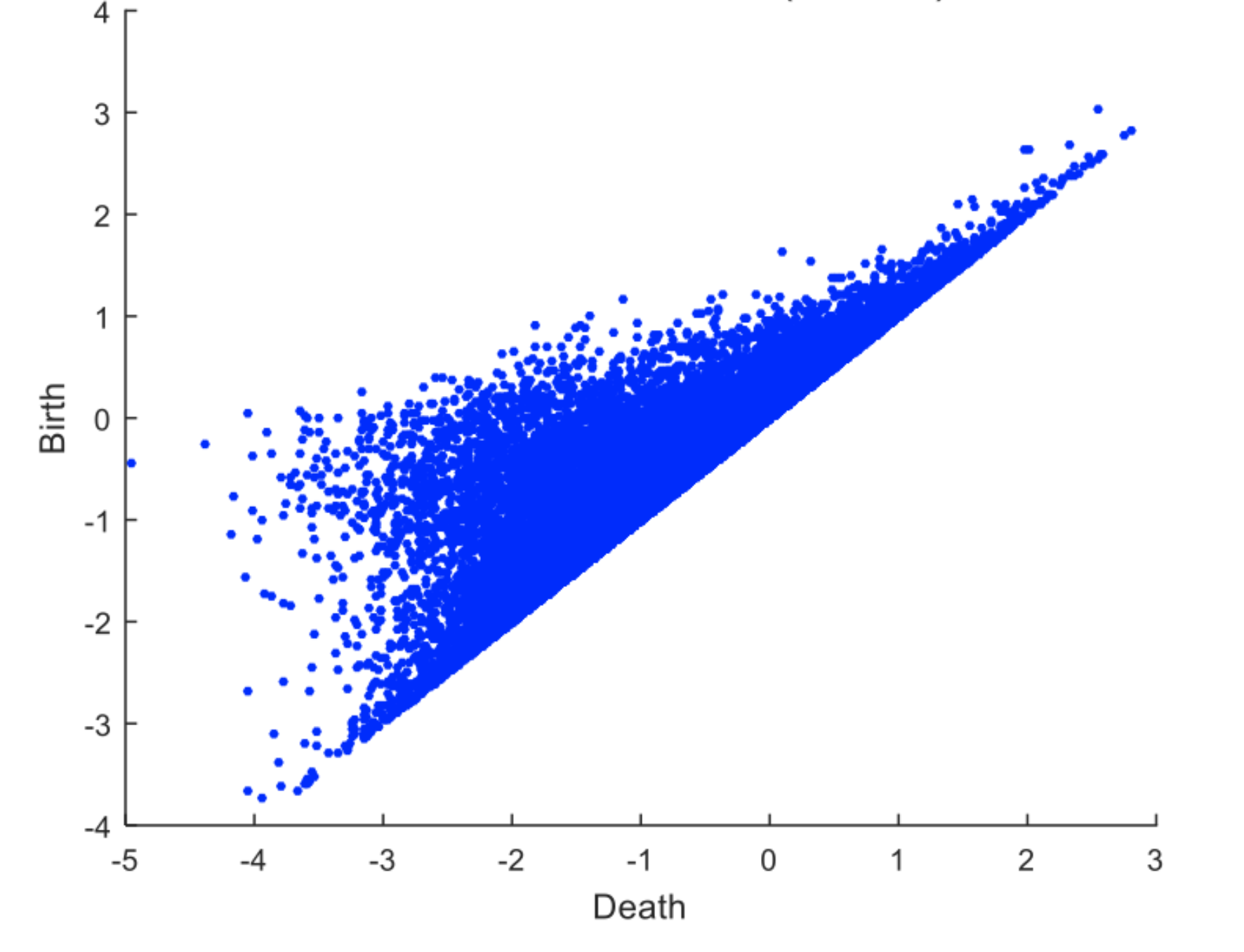}
\ec
\caption{\footnotesize{$H_1$ persistence diagrams for unsmoothed CMB data, northern  cap (left) and southern cap (right). There are approximately 27,000 points in each diagram}.}
\label{fig:CMBH}
\end{figure}

The two PDs of Fig.\ \ref{fig:CMBH} are quite similar, and it is hard to see any obvious differences between them. However, fitting a Gibbs model with pseudolikelihood \eqref{eq:pseudo}, again taking $K=2$, to each of our 32  PD yields some surprising results, summarized in Table \ref{table:cmb}.
\begin{table}[h]
\begin{center}
\begin{tabular}{|c|rrrrrrrr|}
\hline
Smoothing&300&180&120&90&60&40&20&10\\ \hline
$H_0$&0&1&3&3& 3& 3&5 & 1  \\
\hline
$H_1$&0&0&3&2 &2 & 3 & 5 &  5     \\
 \hline
\end{tabular}
\end{center}
\caption{{{\normalfont\sffamily\fontsize{7}{9} {\footnotesize  Number of north-south CMB parameter differences for difference smoothings and homologies.   See  text for details.}}}}
\label{table:cmb}
\end{table}
Each such model involves 5 free parameters (we treat $\delta$ as a nuisance parameter only), and Table \ref{table:cmb} gives the number of such parameters that, for each smoothing, and for each PD ($H_0$ or $H_1$),  were found to be significantly different between the models for the northern and southern caps.  Significance was established using both
the FDR method of \cite{BH} and  the classical Bonferroni correction for multiple tests, both at a 5\% significance level. (Both tests gave  identical results.)

The results are quite striking. At the highest levels of smoothing, there is no  evidence of a difference between the models for the PDs, regardless of homology. At the lower levels, the differences become more and more palpable. While we do not have a clear physical explanation for this, it is most likely due to the effect of interactions between objects that evolved due to the true primordial  CMB fluctuations and  foreground phenomena that evolved at later epochs.

However, whatever is the cosmological reason underlying Table \ref{table:cmb}, the implication is that it is unreasonable to  assume that the  northern and southern cap CMB maps are realizations of the same stochastic process. In other words, an hypothesis of homogeneity is not tenable.

From the point of view of this paper, however, our main discovery is not  cosmological, but lies in demonstrating  the ability of the Gibbs model, {\it which assumes nothing about the original data, nor about how PDs express properties of the  underlying data,} to differentiate between complex structures using purely topological methods. Consequently, we believe that the approach described here will open the door to developing a wide variety of (semi-parametric) statistical methods for further applications of TDA.

\section{Acknowledgments}
{Among others, we thank Jose Blanchet,   Herbert Edelsbrunner, Jingchen Liu, Anthea Monod,
Sayan Mukerjee, and Katherine Turner for helpful conversations in various stages of this research, which was supported in part by SATA II, AFOSR,  FA9550-15-1-0032, and URSAT, ERC  Advanced Grant 320422.}



\bibliographystyle{plain}
\bibliography{paper}

\end{document}